\documentstyle[12pt,fullpage,float,psfig]{article}
\restylefloat{figure}
\restylefloat{table}

\title{A Model for
Isotropic
Crystal Growth from Vapor on a Patterned Substrate}

\author{M. Khenner$^{1}$ ~~R.J. Braun$^{1}$ ~~and M.G. Mauk$^{2}$
\vspace{0.5cm}\\
$^{1}$ Department of Math. Sciences \\
University of Delaware, Newark, DE 19716 \\~\\
$^{2}$ AstroPower, Inc. \\
Solar Park, Newark, DE 19716}

\newcommand{\Section}[1]{\setcounter{equation}{0} \section{#1}}
\newcommand{\rf}[1]{(\ref{#1})}
\newcommand{\beq}[1]{ \begin{equation}\label{#1} }
\newcommand{\eeq}{\end{equation} }

\begin{document}
\pagestyle{plain}
\maketitle

\begin{abstract}

\noindent
We developed a consistent mathematical model for isotropic
crystal growth on a substrate 
covered by the mask material with a periodic series
of parallel long trenches where the substrate is exposed 
to the vapor phase.
Surface diffusion and the flux of particles from vapor are 
assumed to be the main mechanisms of growth.
A geometrical approach to the motion of crystal surface in two dimensions is  
adopted and nonlinear evolution equations are solved by a finite-difference 
method.
The
model allows the direct computation of the crystal surface shape, as well
as the study of the effects due to mask regions of effectively nonzero
thickness.
As in experiments, lateral overgrowth of crystal onto the mask and enhanced growth in the region
near the contact of the crystal and the mask  
is found, as well as the comparable crystal shapes. The growth
rates in vertical and lateral directions are investigated.
\vspace*{0.3cm}

\end{abstract}

\Section{Introduction}

\label{Sec1}

Selective area growth (SAG) of compound semiconductor thin films
by (metalorganic) vapor phase epitaxy on patterned, masked 
substrates 
is a useful technique for fabricating semiconductor devices 
\cite{AOKI}-\cite{THRUSH}.
The masked substrates may also be used as a diagnostic tool; the 
crystals grown in artificial geometries may indicate 
what would be successful or unsuccessful conditions
for producing a desired product.  The masked substrates may also
be used simply for fundamental understanding of the crystal
growth process \cite{THRUSH}-\cite{SAKATA}.

Generally, the method consists of the deposition of an epitaxial
layer on defined areas of the substrate, which are exposed to the
vapor phase via windows lithographically etched through a masking layer.
For semiconductor
device processing the mask is usually a dielectric material such as
SiO$_2$ or Si$_3$N$_4$. In some cases, crystal growth on the mask is not 
desirable but,
in reality, growth is observed there more often than not 
\cite{THRUSH,GIBBON,CORONELL}, \cite{NAM}-\cite{MARX}.
The growth on the mask may occur via lateral overgrowth from the window
in the mask or via nucleation on the mask; the narrower the mask the less
likely is nucleation. 
Following \cite{MITCHELL},
we will denote the lateral overgrowth phenomena as
``epitaxial lateral overgrowth'' (ELOG) in case of growth in narrow windows
($< 10\ \mu \mbox{m}$ in width). We refer to the selective growth
on larger areas (hundreds of microns) as SAG.

In SAG and ELOG the overall growth rate and uniformity of the epitaxial deposit have
been reported to be strongly dependent upon the geometry of masked regions \cite{THRUSH}
as well as the total pressure \cite{KAYSER,SAKATA,HIRUMA,SUDO}. In addition, the deposits are characterized
by an enhanced growth at the perimeter of the windows adjacent to the mask
(excellent photographs of the enhanced growth and the ELOG regions of GaInAs(P) and
InP can be found in \cite{THRUSH}, for example). 
This suggests
that material is supplied to the unmasked areas by surface diffusion along
the mask as well as through direct diffusion from the vapor phase.
It is generally concluded that diffusion lengths on the mask and on the
surface of the epitaxial material such as GaAs, Si or GaN are very short
(of the order of $1\ \mu \mbox{m}$).
Under these conditions, the dominant lateral mass transport mechanism
from the mask regions, for distances greater than a few (say $10$) microns,
is usually attributed to vapor-phase diffusion \cite{GIBBON,MITCHELL,SASAKI},
\cite{CORONELL}-\cite{SAKATA}. 
However, surface diffusion
usually dominates 
\cite{MITCHELL,SASAKI,SAKATA,GREENSPAN,FUJII}
for the growth in narrow windows bounded
by narrow mask regions ($ < 10$ microns) and it is this case (ELOG) that we are interested
in studying.

In our opinion, there exist two types of mathematical models for the 
selective-area
growth. 
Type 1 models operate by solving Laplace's
\cite{THRUSH,KORGEL,ZYBURA} or the
diffusion equation \cite{BOLLEN} for the concentration of species in the 
vapor phase,
subject to appropriate boundary conditions.
The solution yields the concentration profiles in the vapor. Surface
diffusion is ignored in these models.

Type 2 models attempt to account for surface diffusion in several ways.
In \cite{CORONELL}, 
the steady-state diffusion equation for the concentration of species on the 
mask is added
to the Laplace's equation in the vapor, while the diffusion 
on the crystal surface is ignored.
In
\cite{GREENSPAN},
the time-dependent diffusion equation for the concentration of species on 
the crystal
surface is considered, while the diffusion on the mask is ignored. 
In \cite{FUJII},
the steady-state diffusion equation describes
diffusion on both the mask and crystal surfaces, again with appropriate
boundary conditions. 

Neither Type 1 nor Type 2 models allow the 
overgrowth of the crystal on the mask (except \cite{CORONELL}), nor do they 
account for the topography
of the mask (i.e., the mask is considered to be of zero thickness).
They also do not compute the crystal surface shape.
As far as
we know, the anisotropic selective-area growth has not been modeled. 

The goal of this paper is to give a physically and
mathematically consistent description of epitaxial semiconductor crystal 
growth on a masked substrate in a surface diffusion-limited regime.
The model is based on two partial differential equations; one for the crystal 
surface dynamics 
and one for the surface concentration of atoms on the mask. 
The substrate is exposed in the periodic series of parallel long trenches 
etched in the mask. Diffusion in the vapor phase is ignored since we consider
ELOG in this study.
Other simplifying assumptions are:
\begin{itemize}

\item The perturbations caused by selective-area epitaxy are independent
of the orientation of the trenches with respect to 
the crystallographic
axes of the substrate. In other words, in this study we ignore the 
anisotropy of the
crystal growth (the results for the 
anisotropic growth will be reported elsewhere \cite{FUTURE});

\item The diffusion coefficients are constants. It means that we have also
assumed that the temperature is constant over the calculation window. 
The assumption of the constant temperature seems
reasonable since we consider the case of the growth on narrow regions
of the substrate (as explained above);

\item The (possible) interaction between densely arrayed growth regions
is ignored. The numerical methodology we make use of in this study is not 
suitable for the latter problem. Other numerical methods (such as the level set
method \cite{SETYAN-BOOK2}) should be employed to study the crystal surfaces originating
from different growth regions beyond the point of their merging.

\end{itemize}
The model
is designed to allow the direct computation of the crystal surface shape. 
In addition, it allows to study effects due to mask regions of effectively nonzero
thickness. 

The equations and added physical boundary conditions and initial
conditions are nondimensionalized using physical constants
estimated for experiments on ELOG and SAG of GaAs-type materials 
\cite{NISHINAGA}-\cite{NISHINAGA-SHEN}.
A finite difference method is used to find approximate 
solutions to the nondimensionalized, fully nonlinear parametric
evolution equations which govern the crystal surface dynamics.
The numerical solutions exhibit the crystal overgrowth onto the mask
and enhanced growth in the region adjacent to the contact point
(a contact line in three dimensions) where
the mask and the crystal surfaces meet.

The outline of the paper is as follows. In Section \ref{Sec2} we formulate the 
mathematical model, list physical constants, provide details of
nondimensionalization and find an approximate solution to the problem
for diffusion of the concentration on the mask. 
In Section \ref{Sec3} we describe the numerical
method we use to solve the nonlinear initial/boundary value problem
for the crystal/vapor interface. In Section \ref{Sec4} we present the numerical results
for the ELOG in the assumption
of nonzero thickness of the mask. By setting mask thickness to zero 
we obtain more representative results in Section \ref{Sec5}. Finally, 
in  Section \ref{Sec6}
we provide the qualitative comparison of the results obtained to the existing 
results we found in the literature and
give conclusions and future directions for research.

\section{Mathematical model} 

\label{Sec2}

In this section, we describe the mathematical model which is
an initial/boundary value problem consisting of two partial
differential equations, one describing the surface concentration
of the atoms on the mask and the other involving the motion of
the crystal/vapor interface originating from the exposed substrate.
We also introduce physical constants and use them to nondimensionalize
the problem. This model development was partly given in \cite{MPI}.
%
%
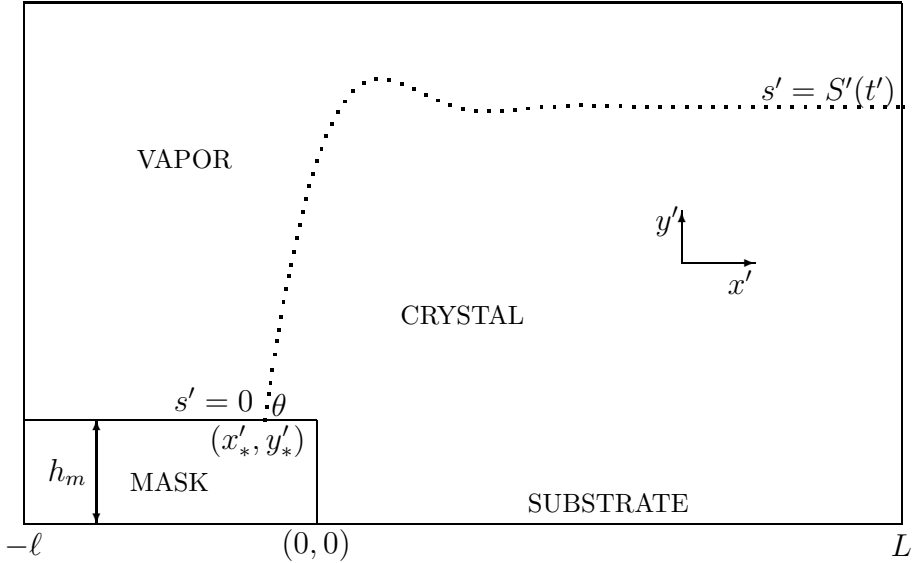
\begin{figure}[H]
\setlength{\unitlength}{0.240900pt}
\ifx\plotpoint\undefined\newsavebox{\plotpoint}\fi
\sbox{\plotpoint}{\rule[-0.200pt]{0.400pt}{0.400pt}}%
\begin{picture}(1500,900)(0,0)
\font\gnuplot=cmr10 at 10pt
\gnuplot
\sbox{\plotpoint}{\rule[-0.200pt]{0.400pt}{0.400pt}}%
\put(60.0,40.0){\rule[-0.200pt]{332.201pt}{0.400pt}}
\put(1439.0,40.0){\rule[-0.200pt]{0.400pt}{197.538pt}}
\put(60.0,860.0){\rule[-0.200pt]{332.201pt}{0.400pt}}
\put(129,122){\makebox(0,0){$h_m$}}
\put(290,106){\makebox(0,0){MASK}}
\put(979,73){\makebox(0,0){SUBSTRATE}}
\put(428,171){\makebox(0,0){$(x'_*,y'_*)$}}
\put(359,237){\makebox(0,0){$s' = 0$}}
\put(462,227){\makebox(0,0){$\theta$}}
\put(750,368){\makebox(0,0){CRYSTAL}}
\put(1324,719){\makebox(0,0){$s' = S'(t')$}}
\put(520,7){\makebox(0,0){$(0,0)$}}
\put(60,7){\makebox(0,0){$-\ell$}}
\put(1439,7){\makebox(0,0){$L$}}
\put(1186,420){\makebox(0,0){$x'$}}
\put(1071,516){\makebox(0,0){$y'$}}
\put(313,614){\makebox(0,0){VAPOR}}
\put(60.0,40.0){\rule[-0.200pt]{0.400pt}{197.538pt}}
\put(175,40){\vector(0,1){164}}
\put(175,204){\vector(0,-1){164}}
\put(1094,450){\vector(0,1){82}}
\put(1094,450){\vector(1,0){115}}
\sbox{\plotpoint}{\rule[-0.500pt]{1.000pt}{1.000pt}}%
\put(438,204){\usebox{\plotpoint}}
\multiput(438,204)(2.836,20.561){2}{\usebox{\plotpoint}}
\put(443.67,245.12){\usebox{\plotpoint}}
\multiput(446,262)(2.836,20.561){2}{\usebox{\plotpoint}}
\put(452.90,306.69){\usebox{\plotpoint}}
\put(456.36,327.15){\usebox{\plotpoint}}
\multiput(459,345)(3.779,20.409){2}{\usebox{\plotpoint}}
\put(466.55,388.57){\usebox{\plotpoint}}
\put(470.20,408.99){\usebox{\plotpoint}}
\put(474.32,429.34){\usebox{\plotpoint}}
\multiput(478,447)(4.409,20.282){2}{\usebox{\plotpoint}}
\put(487.39,490.21){\usebox{\plotpoint}}
\put(491.97,510.46){\usebox{\plotpoint}}
\put(496.73,530.66){\usebox{\plotpoint}}
\put(501.70,550.81){\usebox{\plotpoint}}
\put(506.92,570.90){\usebox{\plotpoint}}
\put(513.22,590.67){\usebox{\plotpoint}}
\put(519.18,610.55){\usebox{\plotpoint}}
\put(526.05,630.13){\usebox{\plotpoint}}
\put(533.40,649.53){\usebox{\plotpoint}}
\put(542.16,668.32){\usebox{\plotpoint}}
\put(551.44,686.89){\usebox{\plotpoint}}
\put(562.71,704.28){\usebox{\plotpoint}}
\put(575.60,720.50){\usebox{\plotpoint}}
\put(592.42,732.61){\usebox{\plotpoint}}
\put(611.94,739.00){\usebox{\plotpoint}}
\put(632.64,738.23){\usebox{\plotpoint}}
\put(652.24,731.92){\usebox{\plotpoint}}
\put(671.27,723.87){\usebox{\plotpoint}}
\put(689.83,714.59){\usebox{\plotpoint}}
\put(708.58,705.71){\usebox{\plotpoint}}
\put(727.82,698.06){\usebox{\plotpoint}}
\put(747.80,692.64){\usebox{\plotpoint}}
\put(768.25,689.13){\usebox{\plotpoint}}
\put(788.99,689.00){\usebox{\plotpoint}}
\put(809.75,689.00){\usebox{\plotpoint}}
\put(830.34,691.00){\usebox{\plotpoint}}
\put(850.91,693.00){\usebox{\plotpoint}}
\put(871.50,695.00){\usebox{\plotpoint}}
\put(892.17,696.03){\usebox{\plotpoint}}
\put(912.85,697.00){\usebox{\plotpoint}}
\put(933.52,698.00){\usebox{\plotpoint}}
\put(954.23,697.46){\usebox{\plotpoint}}
\put(974.95,697.00){\usebox{\plotpoint}}
\put(995.71,697.00){\usebox{\plotpoint}}
\put(1016.38,696.00){\usebox{\plotpoint}}
\put(1037.13,696.00){\usebox{\plotpoint}}
\put(1057.89,696.00){\usebox{\plotpoint}}
\put(1078.65,696.00){\usebox{\plotpoint}}
\put(1099.40,696.00){\usebox{\plotpoint}}
\put(1120.16,696.00){\usebox{\plotpoint}}
\put(1140.91,696.00){\usebox{\plotpoint}}
\put(1161.67,696.00){\usebox{\plotpoint}}
\put(1182.42,696.00){\usebox{\plotpoint}}
\put(1203.18,696.00){\usebox{\plotpoint}}
\put(1223.93,696.00){\usebox{\plotpoint}}
\put(1244.69,696.00){\usebox{\plotpoint}}
\put(1265.44,696.00){\usebox{\plotpoint}}
\put(1286.20,696.00){\usebox{\plotpoint}}
\put(1306.96,696.00){\usebox{\plotpoint}}
\put(1327.71,696.00){\usebox{\plotpoint}}
\put(1348.47,696.00){\usebox{\plotpoint}}
\put(1369.22,696.00){\usebox{\plotpoint}}
\put(1389.98,696.00){\usebox{\plotpoint}}
\put(1410.73,696.00){\usebox{\plotpoint}}
\put(1431.49,696.00){\usebox{\plotpoint}}
\put(1439,696){\usebox{\plotpoint}}
\sbox{\plotpoint}{\rule[-0.200pt]{0.400pt}{0.400pt}}%
\put(60.0,204.0){\rule[-0.200pt]{110.814pt}{0.400pt}}
\put(520.0,40.0){\rule[-0.200pt]{0.400pt}{39.508pt}}
\end{picture}
\vspace*{0.3cm}\\
\parbox{5.0in}{\caption{A sketch of the mathematical situation.  
The free surface of the
growing crystal (curve in two dimensions) is defined parametrically as 
$y'=y'(s',t'),\ x'=x'(s',t'),\ 0 \le s' \le S'(t')$, where $s'$ is the arc length
along the curve and $S'$ is the total arc length of the curve.
The surface
is sketched such that a crystal overgrowth onto the mask
and the region of the enhanced growth (``bump") near the contact point are 
shown.}
\label{Fig1}}
\end{figure}

\subsection{Surface diffusion over the mask}

Our study is on the flat surface covered by the mask material
with a periodic series of parallel long trenches where the 
substrate is exposed.
We examine the growth behavior on a partial cross-section which is a line
segment extending from the center line of one of the masked 
``plateaus" to the center of the next trench; a sketch of the mathematical
situation is given in Figure \ref{Fig1}.
We assume the surface behavior is constant in the perpendicular
direction to legitimize this two-dimensional study.

The interval $-\ell \le x' \le 0$ corresponds to the zone where the 
thin mask is located.
Here we study the surface diffusion of the concentration 
$n_m = n_m(x^\prime, t^\prime)$ of atoms on the 
horizontal part of the mask not yet occupied by the growing crystal:
\begin{equation}
{\partial n_m \over \partial t^\prime} = D_s^{(m)} {\partial^2 n_m
\over \partial (x^{\prime})^2 } + J_g - {n_m \over \tau_m},\quad
-\ell \le x' \le x'_*(t')
\label{2.1}
\end{equation}
where, $J_g$ represents uniform flux of atoms from a
vapor and impinging on the mask.
The term $n_m /\tau_m$ provides the concentration of atoms leaving
the surface to go into the vapor.
The constant $\tau_m$ is the mean residence time
on the step.
Finally, $D_s^{(m)}$ is the surface diffusivity of the 
atomic concentration on the mask.
Note that \rf{2.1} has been given in \cite{NISHINAGA,CHERNOV},
for example.

Appended to \rf{2.1} is a symmetry boundary condition,
\begin{equation}
{\partial n_m \over \partial x^\prime}(-\ell, t') = 0.
\label{2.2}
\end{equation}
We assume that the atoms close to the crystal will be absorbed
quickly; thus we take the boundary condition at the contact point 
to be a perfect sink:
\begin{equation}
n_m(x'_*(t'), t') = 0.
\label{2.3}
\end{equation}
An initial condition is also needed to define this part of the problem
completely, say
\begin{equation}
n_m(x', 0) = 0,\quad -\ell \le x' \le 0.
\label{2.4}
\end{equation}

To summarize, there exist two options for every atom arriving to the horizontal 
part of the {\it macroscopic}
step which represents the mask at $-\ell \le x' \le 0$: it can either desorb
back into the vapor phase or be transported by surface diffusion
from left to right.
The drifting atoms are incorporated in the crystal at the contact point
$(x'_*,y'_*)$ as they meet the crystal on the way. At the initial stage
of the growth the atoms need to diffuse
over the mask corner prior to being incorporated in the crystal.
We don't study surface diffusion
along the vertical part of the step and assume that atoms diffused
over the corner are incorporated
in the growing crystal instantly. This approximation is good if
the mask thickness is much less than the mask region width, that is if
$h_m \ll \ell$ (in ELOG the ratio $h_m/\ell$ is usually $< 0.1$ 
\cite{THRUSH,GIBBON,SASAKI}, \cite{SAKATA}-\cite{HIRUMA},
\cite{GREENSPAN}. 
In SAG
this ratio is so small that the mask can be considered to be of vanishing
thickness). Both parts of the mask are assumed to be physically
equivalent. As a particular consequence, the constant equilibrium angle 
$\theta(\gamma,\gamma_m)$
which the crystal surface forms with the mask
at the contact point (as a result of the assumed thermodynamic equilibria
in the vicinity of the junction \cite{HERRING}; $\gamma$ and $\gamma_m$
are the surface energies of the crystal and of the mask respectively) should be the same
and independent of the contact point position on the step, 
ref. Figure \ref{Fig1}.

\subsection{Model for the crystal surface}

We now turn to the formulation of the equations and the associated boundary
and initial conditions for the motion of the interface between the crystal
that is built-up from the substrate on the interval $0 \le x' \le L$ and the
vapor.  The interface is allowed to move  above the mask
on the interval $-\ell \le x' < 0$. This development of the model closely 
parallels that of Mullins \cite{MULLINS57} in his now classic study
of grain boundary grooving.

The interface moves via two mechanisms.  The first is from a flux of atoms 
to the interface from the vapor; this flux is normal to the interface.
This contribution to the normal velocity of the interface is proportional 
to the jump in 
the chemical potential across the interface, viz.,
\begin{equation}
M A (\mu_V^T - \mu_c).
\label{2.5}
\end{equation}   
Here $M$ is the mobility of the interface and $A=\partial n_s/ \partial \mu$
is the change of surface concentration $n_s$ with chemical potential $\mu$;
$\mu_c$ is chemical potential of the crystal surface and is given by
\begin{equation}
\mu_c = \mu_\infty + \Omega \gamma \kappa,
\label{2.6}
\end{equation}
where $\gamma$ is the surface energy of the crystal, $\Omega$ is the atomic
volume, $\kappa$ is the curvature of the crystal surface and 
$\mu_\infty$ is the reference value of the chemical potential for 
the crystal with a flat surface.
$\mu_V^T$ is the chemical potential of the vapor and it may be written
as $\mu_V^T = \mu_V + \mu_\infty$.  
Comparison with \cite{MULLINS57} reveals that $A$ plays a 
role identical to $n_s/kT$, 
where $n_s$ is the surface concentration, $k$
is Boltzmann's constant, and $T$ is the absolute temperature.

The second contribution is from the surface diffusion of atoms along 
the interface; the current of atoms $J_s$
is proportional to the gradient of the chemical potential along the
surface.  For the curve in the plane, the surface gradient is the derivative
with respect to arc length, hence 
\begin{equation}
J_s = - A D_s^{(c)} {\partial \mu_c \over \partial s^\prime}.
\label{2.7}
\end{equation}
The contribution to the rate of increase in the atoms per unit area
(proportional to the normal velocity) is proportional to 
the surface divergence of this flux; for the curve,
we have $ (-\partial J_s/\partial s')$.

The net rate of increase in the number of atoms per unit area
(or the time rate of change of the concentration) may be written
as $V'_n/\Omega$, where the $\Omega$ is the atomic volume and $V'_n$ is
the normal velocity.  Combining the normal and surface diffusion contributions
results in 
\begin{equation}
V'_n  = A \Omega \left\{ {\partial \over \partial {s^\prime}} \left[ 
D_s^{(c)}
{\partial \over \partial {s^\prime}} ( \Omega \gamma 
\kappa) \right] + M (\mu_V- \Omega \gamma
 \kappa ) \right\}.
\label{2.8}
\end{equation}
The parametric evolution equations for the crystal surface are given
in the Section \ref{evoleq}.
Next, we describe physical boundary and initial conditions which will be used
for the computation.

First, we assume the following obvious conditions hold for the contact
point location $x'_*, y'_*$
(see Figure \ref{Fig1}):
\begin{equation}
x'_* = 0,\quad \mbox{if}\quad 0 \le y'_* < h_m,
\label{2.9}
\end{equation}
$$
y'_* = h_m,\quad \mbox{if}\quad -\ell \le x'_* < 0.
$$
These conditions place the contact point on the mask.

At $x^\prime = L$ we assume the symmetry conditions,
\begin{equation}
{\partial y' \over \partial x^\prime}(L, t') = 0 
\;\;\mbox{and}\;\;
{\partial^3 y' \over \partial (x^\prime)^3}(L, t') = 0.
\label{2.10}
\end{equation}
In \rf{2.10}, the second condition is effectively the condition of no
atomic flux at $x=L$ \cite{MULLINS57}.

Unless otherwise noted, the crystal surface is initially a horizontal line even 
with the substrate,
\begin{equation}
y'(x',0) = 0,\quad 0 \le x' \le L .
\label{2.11}
\end{equation}

After the short time the crystal rearranges to form an equilibrium
angle $\theta$ with the mask at the contact point:
\begin{equation}
t'>0:\quad {\partial y' \over \partial x'}|_{(x'_*,y'_*)} = \tan{\theta},
\quad \mbox{if}\quad -\ell \le x'_* < 0\;\; \mbox{and}\;\; y'_* = h_m,
\label{2.12}
\end{equation}
$$
\quad \quad {\partial y' \over \partial x'}|_{(x'_*,y'_*)} = \tan{(\theta-90^\circ)},
\quad \mbox{if}\quad x'_* = 0\;\; \mbox{and}\;\; 0 \le y'_* < h_m.
$$

We also match the fluxes from the atoms on the mask surface
moving from left-to-right with the flux onto the growing crystal
by requiring that
\begin{equation}
D_s^{(m)} {\partial n_m \over \partial x^\prime }(x'_*,t')
= A D_s^{(c)} \Omega \gamma 
{\partial \kappa \over \partial {s^\prime} }(x'_*,t').
\label{2.13}
\end{equation}

\subsubsection{Evolution equations for the crystal surface}

\label{evoleq}

We take a geometrical approach to the problem of the motion of an
open curve (crystal surface) $\gamma(t')$ along it's
normal vector field with
a speed function of curvature and some of it's derivatives.

Let ${\mathbf x}'(r,t')$ be the position vector which, at time $t'$, parametrizes
$\gamma(t')$ by $r,\ 0 \le r \le R(t')$ ($r$ is not necessarily the arc length). 
With $\kappa(r,t')$ as the curvature at 
${\mathbf x}'(r,t')$,
the equations of motion are
\begin{equation}
\label{2.14}
{\mathbf n}(r,t')\cdot\frac{\partial {\mathbf x}'(r,t')}{\partial t'} = 
V'_n(\kappa(r,t'),\kappa_r,...),
\end{equation}
$$
{\mathbf x}'(r,0) = \gamma(0)\quad \mbox{prescribed},
$$
where ${\mathbf n}(r,t')= -{\mathbf i}\ \partial y'/\partial r + 
{\mathbf j}\ \partial x'/\partial r$ is the unit normal vector at ${\mathbf x}'(r,t')$
and ${\mathbf i}$ and ${\mathbf j}$ are the unit vectors along $x$ and $y-$axis
respectively.

Written in terms of the coordinates ${\mathbf x}'(r,t') = (x'(r,t'),y'(r,t'))$, an
equivalent formulation is \cite{Sethian_CMP}
\begin{equation}
\label{2.15}
\frac{\partial x'}{\partial t'} = V'_n\ \frac{1}{g'}\ \frac{\partial y'}{\partial r},
\end{equation}
$$
\frac{\partial y'}{\partial t'} = -V'_n\ \frac{1}{g'}\ \frac{\partial x'}{\partial r},
$$
where 
\begin{equation}
g' = \sqrt{\left(\frac{\partial x'}{\partial r}\right)^2+ 
\left(\frac{\partial y'}{\partial r}\right)^2} 
\label{2.16}
\end{equation}    
is metric function (the metric function measures the ``stretch" of the
parametrization). The relation 
\begin{equation}
\label{2.17}
ds' = g'(r,t)\ dr
\end{equation}
can be used to cast the normal velocity \rf{2.8} in terms of $r$.
The expression for the curvature in terms of $x',y'$ is
\begin{equation}
\label{2.18}
\kappa = \frac{1}{(g')^3}\left(
\frac{\partial^2 y'}{\partial r^2}\ \frac{\partial x'}{\partial r} -
\frac{\partial^2 x'}{\partial r^2}\ \frac{\partial y'}{\partial r}\right).
\end{equation}

\subsection{Physical parameters}

The physical constants listed in Table \ref{Table1} are representative for ELOG of
GaAs-type materials 
at temperatures near $650\ K$.
{\small
\begin{table}
\caption{Physical constants.}
\label{Table1}
\begin{center}
\begin{tabular}{|c|c|c|} \hline
Constant & Description & Value/Units \\ \hline\hline
$J_g$ & Atomic flux from vapor
  & $10^{15}$ Atoms/($\mbox{cm}^2\cdot$sec)\\
$\tau_m$ & Mean residence time of atoms on mask & 1 sec\\
$D_s^{(m)}$ & Diffusivity on mask
 & $5 \times 10^{-8}$ $\mbox{cm}^2/\mbox{sec}$\\
$D_s^{(c)}$ & Diffusivity on crystal surface
 & $2\times 10^{-8}$ $\mbox{cm}^2/\mbox{sec}$\\
$\Omega$ & Atomic Volume & $2 \times 10^{-23}$ $\mbox{cm}^3/\mbox{Atom}$\\
$\gamma$ & Surface energy & $10^3$ $\mbox{ergs}/\mbox{cm}^2$ \\
$A$ & Change in concentration/Change in chemical potential
 & $2 \times 10^{28}$ $\mbox{Atoms}^2$/(erg$\cdot \mbox{cm}^2$)\\
$\mu_V$ & Chemical potential in the vapor
 & $3 \times 10^{-13}$ $\mbox{erg}\cdot \mbox{atom}$ \\
$M$ & Mobility & $10\ \mbox{sec}^{-1}$\\
\hline
\end{tabular}
\end{center}
\end{table}
}

Since the parameter space is large, the parameters in Table \ref{Table1} were 
fixed in the course of our study of ELOG. We only investigated the influence
of the four geometrical parameters, $L, \ell, h_m$ and $\theta$. These 
parameters (except $\theta$) can be controlled in the experiment given the 
epitaxial material and
growth conditions; changing materials changes $\theta$. We varied $L, \ell, h_m$ and $\theta$ within the 
intervals listed in Table \ref{Table2}.
{\small
\begin{table}
\caption{Geometrical parameters.}
\label{Table2}
\begin{center}
\begin{tabular}{|c|c|c|c|} \hline
Parameter & Description & Value/Units \\ \hline\hline
$L$ &  Width of substrate region & $10^{-4} - 10^{-3}$ $\mbox{cm}\ (1 - 10\ \mu \mbox{m})$ \\
$\ell$ & Width of mask region & $10^{-4} - 10^{-3}$ $\mbox{cm}\ (1 - 10\ \mu \mbox{m})$ \\
$h_m$ & Height of mask above substrate & $0 - 1.25\times 10^{-4}$ $\mbox{cm}\ (0 - 1.25\ \mu \mbox{m})$ \\
$\theta$ & Contact angle & $0^\circ - 180^\circ$  \\
\hline
\end{tabular}
\end{center}
\end{table}
}

\subsection{Nondimensionalization and diffusion problem over the mask}

To complete this section we now describe the nondimensionalization used
and solve the problem for the concentration on the mask.
Table \ref{Table3} has all the nondimensional parameters.
\begin{table}
\caption{Nondimensional parameters.}
\label{Table3}
\begin{center}
\begin{tabular}{|c|c|c|} \hline
Constant & Expression & Approximate Value (for $L=\ell=5\ \mu \mbox{m}$)\\ \hline\hline
$\epsilon$ & $D_s^{(c)}/D_s^{(m)}$ & $4 \times 10^{-1}$ \\
$D$ & $(\Omega^2 A \gamma)/(L^2)$ & $3\times 10^{-8}$ \\
$\delta$ &  $(M \Omega^2 \gamma A)/ D_s^{(m)}$ & $1.6\times 10^{-6}$ \\
$\alpha$ & $L^2/(\tau_m D_s^{(m)}$) & $5$ \\
$J$ & $ (\Omega A \mu_V M L)/ D_s^{(m)}$ & $1.2\times 10^{-2}$ \\
$\bar h_m$ & $h_m/L$ &  $10^{-1}$ \\
$\beta$ & $(A \Omega \gamma) / (\tau_m J_g L)$ & $8 \times 10^{-4}$ \\
$d$ & $\ell/L$ & 1 \\
$f_m$ & $ \sqrt{\alpha}\tanh (\sqrt{\alpha} d)$ & 2.18 \\
\hline
\end{tabular}
\end{center}
\end{table}

As above, we first handle the diffusion problem over the mask. 
Letting
\[
x^\prime = L x, \;\;\; t^\prime = {L^2 \over D_s^{(m)} } t,
\;\;\mbox{and}\;\; n_m = \tau_m J_g u
\]
we can rewrite \rf{2.1}-\rf{2.4} as 
\begin{equation}
{\partial u \over \partial t}
 = {\partial^2 u \over \partial x^2}+\alpha (1 - u),\quad
-d \le x \le x_*(t)
\label{2.19}
\end{equation}
with boundary conditions
\begin{equation}
{\partial u \over \partial x}(-d,t) = 0\;\;\mbox{and}\;\; u(x_*(t), t)=0
\label{2.20}
\end{equation}
and an initial condition
\begin{equation}
u(x, 0)=0.
\label{2.21}
\end{equation}

When the contact point position is fixed at $x_*(t)=0$, an exact solution
to the problem \rf{2.19} - \rf{2.21} may be found \cite{MPI}. In that case,
a constant flux may be found from the steady state solution to \rf{2.19} (with
$x_*(t)=0$).
This constant flux is given by
\begin{equation}
\frac{\partial u_{steady}}{\partial x}(x) = - \sqrt{\alpha}\left(\sinh (\sqrt{\alpha} x) +
\tanh (\sqrt{\alpha} d) \cosh (\sqrt{\alpha} x)\right);
\label{2.25} 
\end{equation}
At $x=0$ we have
\begin{equation}
\frac{\partial u_{steady}}{\partial x}(0) = - \sqrt{\alpha}\tanh (\sqrt{\alpha} d).
\label{2.26} 
\end{equation}
The value of $f_m\equiv |\frac{\partial u_{steady}}{\partial x}(0)|$ is
estimated
to be 2.18 for the parameters in Table \ref{Table3}. When the contact
point moves, the same steady flux is not available; however, we will assume
a constant flux $f_m$ from the mask at the contact point for the rest of this 
paper. The flux $f_m$ will be used as a boundary condition for the evolution equations 
of the growing crystal.
The approximation just described is validated in \cite{FUTURE},
where we numerically solve the full surface diffusion problem on the mask and
demonstrate that the contact point indeed propagates steadily along the mask.

We also can estimate the effect of neglecting surface diffusion along 
the vertical part of the step. If we 
replace $d$ with $d+\bar h_m$ in the 
expression \rf{2.26} for the steady flux at the contact point, 
the resulting change in $f_m$ will be very small since
$\bar h_m \ll d$ for ELOG (ref. Table \ref{Table3}).

%
%
%
%
%
%

To nondimensionalize \rf{2.8}-\rf{2.13}, \rf{2.15}-\rf{2.18} we make the following
variable changes, again referring to Table \ref{Table3}
for the definitions of the nondimensional parameters:
\[
y' = L y,\ x' = L x,\ s' = L s,\ g' = L g,\ h_m = L \bar h_m,\ 
\kappa = L^{-1} K \;\;\mbox{and}\;\; t^\prime = {L^2 \over D_s^{(m)} } t. 
\]
The evolution equations then become
\begin{equation}
\label{2.27}
\frac{\partial x}{\partial t} = V_n\ \frac{1}{g}\ \frac{\partial y}{\partial r},
\end{equation}
$$
\frac{\partial y}{\partial t} = -V_n\ \frac{1}{g}\ \frac{\partial x}{\partial r},
$$
where the normal velocity is given by
\begin{equation}
V_n = \epsilon D \frac{1}{g}\frac{\partial}{\partial r}\left(\frac{1}{g}\ 
\frac{\partial K}{\partial r}\right) + J - \delta K
\label{2.28}
\end{equation}
and the metric function and the curvature by
\begin{equation}
g = \sqrt{\left(\frac{\partial x}{\partial r}\right)^2+
\left(\frac{\partial y}{\partial r}\right)^2},\quad 
K = \frac{1}{g^3}\left(
\frac{\partial^2 y}{\partial r^2}\ \frac{\partial x}{\partial r} -
\frac{\partial^2 x}{\partial r^2}\ \frac{\partial y}{\partial r}\right).
\label{2.29}
\end{equation}

We now have that $x=1$ corresponds to $R(t)$ and the
contact point $(x_*,y_*)$ corresponds to $r=0$, see Figure \ref{Fig1}.
The boundary conditions are:
\begin{equation}
{\partial y \over \partial r}(r=R(t), t) = 0,\;\;\;\;
{\partial^3 y \over \partial r^3}(r=R(t), t) = 0,
\label{2.30}
\end{equation}
\begin{equation}
\frac{\partial y/ \partial r}{\partial x/ \partial r}(r=0, t) = \tan{\theta},
\quad \mbox{if}\quad -d \le x_* < 0\;\; \mbox{and}\;\; y_* = \bar h_m,
\label{2.31}
\end{equation}
$$
\frac{\partial y/ \partial r}{\partial x/ \partial r}(r=0, t) = \tan{(\theta-90^\circ)},
\quad \mbox{if}\quad x_* = 0\;\; \mbox{and}\;\; 0 \le y_* < \bar h_m.
$$
and 
\begin{equation}
\frac{1}{g}{\partial K \over \partial r}(r=0, t)
 = \frac{f_m}{\epsilon \beta}.
\label{2.32}
\end{equation}
The initial condition is
\begin{equation}
y(r, 0) = 0,\ x(r, 0) = r,\ 0 \le r \le 1.
\label{2.33}
\end{equation}
And, the nondimensional coordinates of the contact point satisfy
\begin{equation}
x_* = 0,\quad \mbox{if}\quad 0 \le y_* < \bar h_m,
\label{2.34}
\end{equation}
$$
y_* = \bar h_m,\quad \mbox{if}\quad -d \le x_* < 0.
$$

\Section{Numerical method}

\label{Sec3}

Evolution of the crystal surface is solved by a finite-difference method.
$N$ marker particles are placed along the surface and 
advanced according to equations \rf{2.27}.
Centered,
second order accurate and conservative finite difference approximations are used for 
the spatial
derivatives in \rf{2.27}; one-sided
finite differences are used where needed to compute derivatives at the
boundaries $r=0$ and $r=R(t)$. This yields a set of coupled ordinary differential
equations for the motion of marker particles. The time derivatives are
approximated by the forward Euler method.  

The computational cycle consists of the following steps:

\begin{itemize}

\item Given the coordinates $(x,y)$ of marker particles at time $t$, compute
$g,\ K$ (including $g_0$ and $K_0$), the derivatives of curvature
and then $V_n$ in the interior points. $K_0$ is computed
by imposing the boundary condition \rf{2.32};

\item Using Euler scheme, compute coordinates $(x,y)$ of the 
interior marker particles at time $t+\Delta t$, where $\Delta t$ is
the time step;

\item Taking into account \rf{2.34} and using 2nd order accurate 
one-sided finite differences,
compute coordinates $x_*,y_*$ of the contact point at time $t+\Delta t$ 
from \rf{2.31};

\item Remesh, i.e. redistribute the marker particles along the surface in order to
preserve uniform spacing.
With this purpose, the parametric cubic spline representation
of the curve is constructed. These splines are then used to redistribute
the marker particles at evenly spaced arc lengths (also, the code has an option
to dynamically increase the number of markers along the surface once
it becomes too elongated). 

\end{itemize}

\begin{figure}[H]
\centering
\psfig{figure=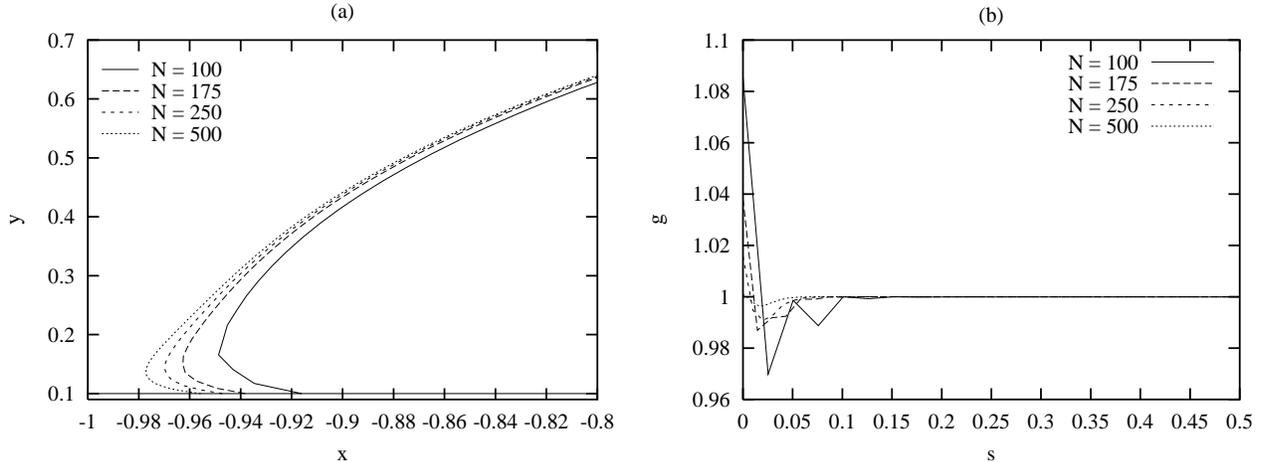,height=2.4in,width=6.5in,angle=270}
\caption{(a): Crystal surface at $t=90$, computed with different
number of marker particles (ref. Figure \ref{Fig2}(b) for the values of
parameters); (b): Corresponding metric functions $g(s)$.}
\label{Fig_rob}
\end{figure}

The remeshing justifies the proper choice of the parametrization -
it ensures the parameter is the arc length, $r=s$. Therefore, $r$ lies
in the time-dependent interval $\left[0,S(t)\right]$, where $S(t)$ is
the total arc length of the curve.
The remeshing procedure is computationally expensive but it ensures
the smoothness of the surface for 
the entire simulation period.
With this procedure, almost no ``stretch" of the parametrization occurs and
the metric function becomes unity everywhere in $\left[0,S(t)\right]$,
except the very narrow region near $s=0$ where the boundary pollutes
the data. However, this error decreases rapidly with the grid refinement
(see Figure \ref{Fig_rob}(b)).
The surface rearranges quickly at the junction point to satisfy the 
equilibrium contact angle $\theta$ with the mask at the contact point.

\begin{figure}[h]
\centering
\psfig{figure=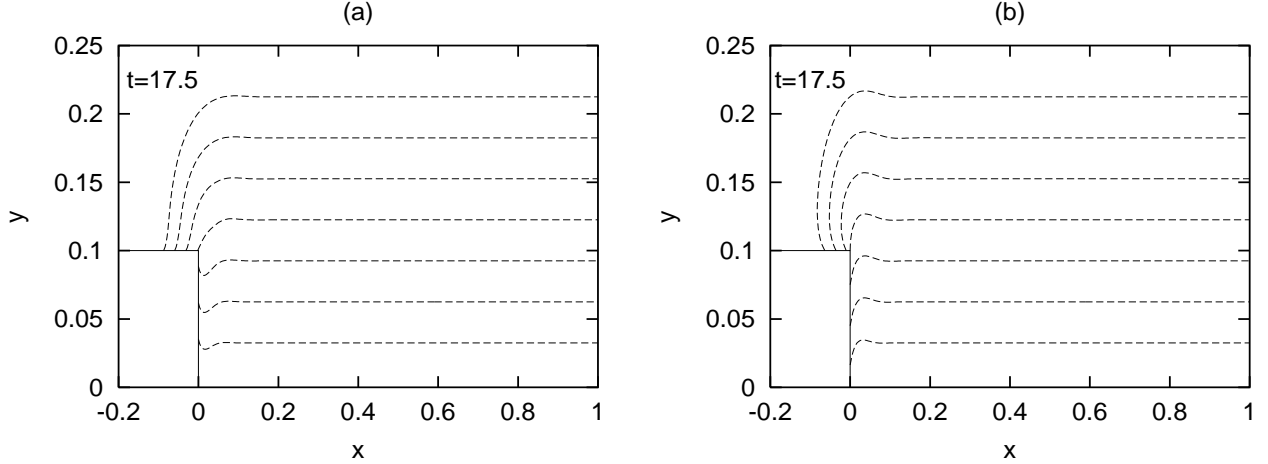,height=2.4in,width=6.5in,angle=270}
\caption{The evolving crystal surface.
(a): $\theta=30^\circ$, (b): $\theta=150^\circ$.
$L=\ell=5\ \mu \mbox{m},\ h_m = 0.5\ \mu \mbox{m}$.
The surface profiles are dumped every 2500 time
steps.
The time labels correspond to the time
at which the last profile is dumped. In (a) and (b), $\theta$ is the
angle between the mask and the interior of the crystal at the contact point.
The profiles shown at $t=10$ (fourth from the bottom in (a) and (b)) 
have
the contact point exactly on the mask corner where $\theta$ readjusts, ref.
the discussion below.}
\label{Fig2}
\end{figure}

To proove the validity of the numerical code, we made a series of runs 
with $N = 100, 175, 250, 500$ and, in Figure \ref{Fig_rob} plotted the 
surface profiles near the contact point and the corresponding metric 
functions (at $t=90$, 
when the overgrowth on the mask is almost completed) for the case of
Figure \ref{Fig2}(b). The inspection of Figure \ref{Fig_rob} suggests
that $N=250$ marker particles provide accurate enough resolution of 
the curve; the difference in the $x$-coordinates of the contact point 
for $N=250$ and $N=500$ is $0.0085$, and the deviation of $g(0)$ from
unity for $N=250$ is $1.6\%$.
Thus, the computations presented in the Sections \ref{Sec4} and \ref{Sec5} 
were done with $N=250$ marker particles. 
We verified all the results  
by performing runs with $N=500$. 

As a further check on the numerical code, we set $\delta = J = 0$ in 
\rf{2.28}, $f_m = 0$ in \rf{2.32}, $\bar h_m=0$ in
\rf{2.31}, \rf{2.34} and computed (for different contact angles $\theta$) 
the evolution of a retracting solid film step.
We obtained the same surface profiles as in \cite{MIKSIS}, where the latter problem is
considered in detail (comparison not shown).

\Section{Numerical results for mask layer of nonzero thickness}

\label{Sec4}

As an example of the crystal surface evolution, Figure \ref{Fig2} shows 
the surface profiles for the contact angles
$\theta=30^\circ$ and $\theta=150^\circ$ at different values of the nondimensional
time (note the different scales along $x$ and $y$-axis).

\begin{figure}[h]
\centering
\psfig{figure=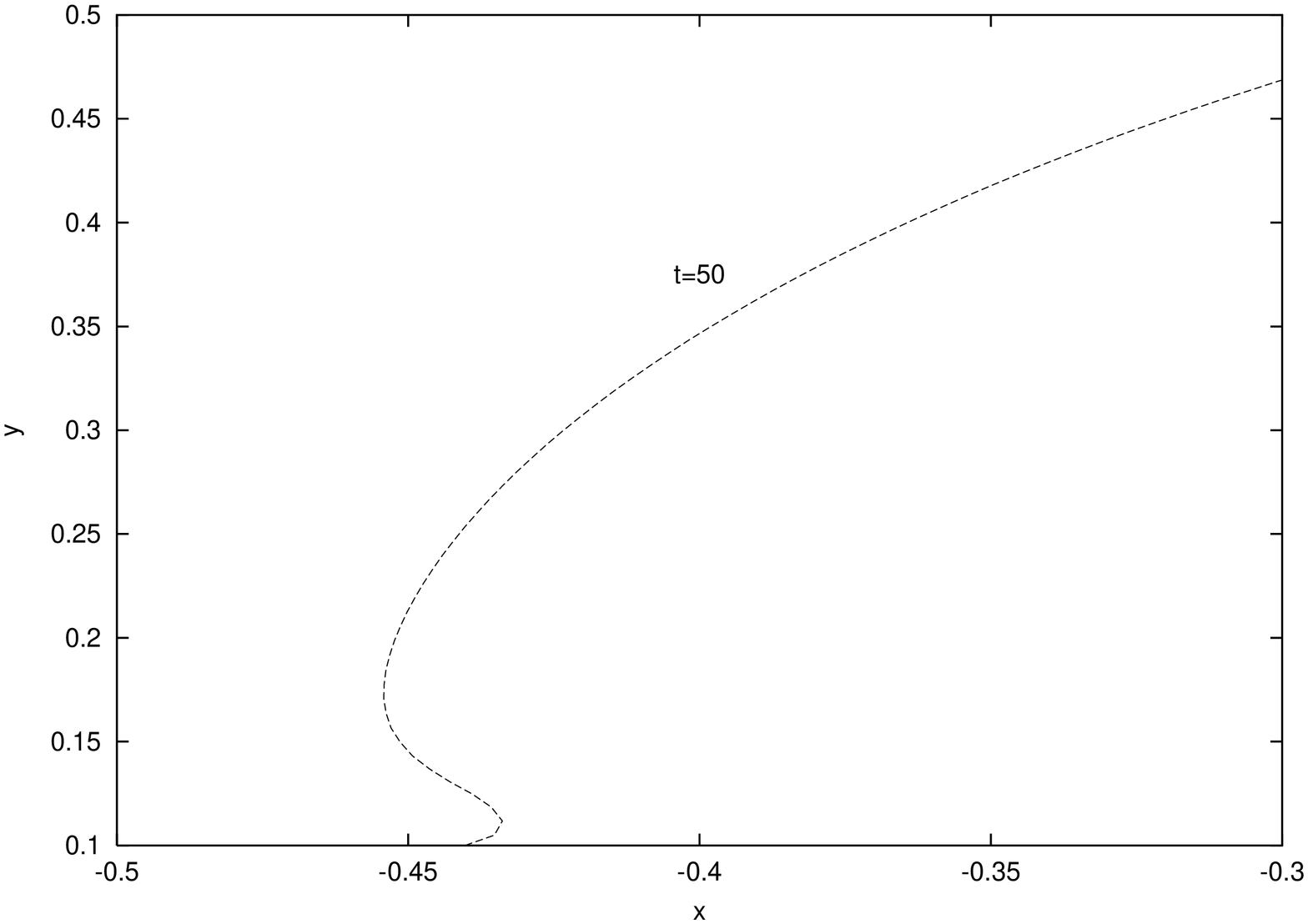,height=2.5in,width=4.5in,angle=270}
\parbox{4.5in}{\caption{Snapshot of the crystal surface shown in Figure \ref{Fig2}(a)
at $t=50$.}
\label{Fig3}}
\end{figure}

\noindent
At later
times, the curve in  Figure \ref{Fig2}(a) bends in the vicinity
of the contact point, see Figure \ref{Fig3}. This indicates that the crystal
prefers not to ``wet" the mask, i.e. the contact angles larger than $90^\circ$
are preferred (as in Figure \ref{Fig2}(b)).
The contact angle is held constant
along the mask surface during the computation except at the corner of the
mask; for further detail see below.

Some comments should be made about the initial stage of
the evolution from zero-level initial condition \rf{2.33} and
about the numerical treatment of the curve
propagation over the step corner. 
First, note that initial condition \rf{2.33} does not match the boundary
condition \rf{2.31}. This implies that a singularity exists at $(x=0,y=0)$
at $t=0$.
This singularity does not present a barrier in solving the system
numerically.
Physically, the
equilibrium angle
should be formed instantaneously compared to the time needed for the
evolution of
the surface. We are not concerned with the details of this instance.
Our numerical experiments reveal that for some values of nondimensional 
parameters
the contact point initially retracts in the unphysical domain 
$y<0$ (below the substrate, case $\bar h_m > 0$) or in the domain $x>0$
(on the substrate, case $\bar h_m = 0$). If this happens we replace the 
initial curve \rf{2.33} with
the hyperbolic tangent curve having the desired contact angle $\theta$ 
with the mask at the 
contact point. For instance, in the case of Figure \ref{Fig2}(a) this
curve may be
\begin{equation}
y(r, 0) = y_0\left( 1 - \tanh{\left(r\frac{\tan{\theta}}{y_0}\right)} \right),\ x(r, 0) = r,\ 0 \le r \le 1
\label{4.1}
\end{equation}
where, $y_0$ gives the small initial elevation of the surface at $x=0$.
Similarly, in the case of Figure \ref{Fig2}(b) the
initial condition may be
\begin{equation}
y(r, 0) = y_0\tanh{\left(r\frac{\tan{\theta}}{y_0}\right)},\ x(r, 0) = r,\ 0 \le r \le 1
\label{4.2}
\end{equation}
where, $y_0$ gives the small initial elevation of the surface at $x=1$.

Similar reasoning is applied to the step corner point.
Numerically, the corner point is singular
in the sense that the contact angle is not well defined there.
One way to deal with this situation is to instantly readjust 
the angle the curve forms with the upper mask surface 
to the equilibrium value once the contact point reaches the
step corner. 
However, such instant readjustment
causes a short retraction into the unphysical domain $(x>0,y=\bar h_m)$.
To avoid this retraction, we pin the contact point
at the corner and
allow the contact angle with respect to the upper mask surface to
readjust to the equilibrium contact angle there. 
Once the desired  angle is reached, the contact point resumes motion along
the mask.
The time interval needed for the 
readjustment is vanishingly small compared to the characteristic times
of the problem. 
These times are as follows.
\begin{itemize}

\item $T_1 =\ $the time needed for the ``bump" on the surface
to disappear. $T_1$ is an important quantity since in 
semiconductor device
manufacture one may not want to grow the crystal with deformed
surface; also, $T_1$ is interesting since it provides some measure of  
surface diffusion's effect on the evolution.

\item $T_2 =\ $the time needed for the contact point to reach the point
$(x=-d,y=\bar h_m)$ on the mask i.e., reach the end of the mask region.

\item $T_3 =\ $the time needed for the contact point to reach the step corner
(only in the case $h_m > 0$).

\end{itemize}

We also define:
\begin{itemize}

\item the normalized contact point position as the ratio of the 
distance
(along the $x$-axis, 
from the contact point to the origin) to the mask region
width, i.e. $|x_*|/d$;

\item the normalized maximum growth thickness as the ratio of the 
maximal elevation of the surface (the ``bump" height)
to the height of the crystal at $x = 1$.

\end{itemize}

\begin{figure}[h]
\centering
\psfig{figure=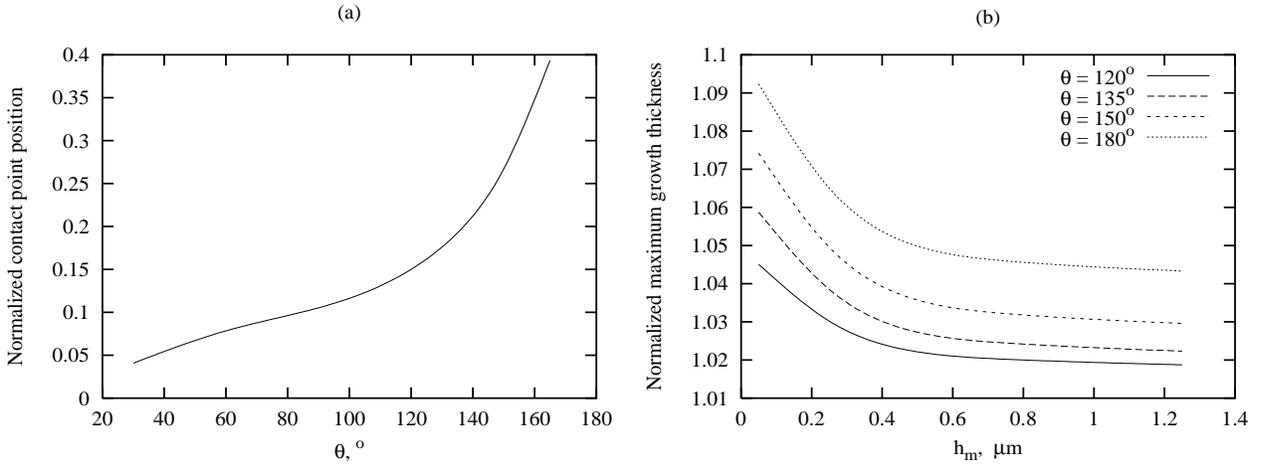,height=2.4in,width=6.5in,angle=270}
\caption{
(a): Normalized contact point position \textit{vs.}\
contact angle at $t = T_1$. 
$L=\ell=5\ \mu \mbox{m},\ h_m = 0.5\ \mu \mbox{m}$.
(b): Normalized maximum growth thickness \textit{vs.}\
mask thickness at $t = T_3$ for different contact angles; $L,\ell$ are
the same as in (a).}
\label{Fig4}
\end{figure}

Figure \ref{Fig4}(a) shows the normalized contact point position \textit{vs.}\
contact angle at $t = T_1$. The crystal overgrowth
onto the mask increases as the contact angle increases (compare to 
Figure \ref{Fig2}(a),(b) where, at the same times, the overgrowth is 
approximately the same but the ``bump" is more pronounced
for $\theta = 150^\circ$, therefore resulting in larger overgrowth
at $t = T_1$). The value of $T_1$ depends on the relative tolerance
chosen to measure the disappearence of the ``bump"; for instance, 
with the relative tolerance
0.005 used to compute the curve in Figure \ref{Fig4}(a) and for 
$\theta=150^\circ$ (case of Figure \ref{Fig2}(b)) the value for 
$T_1$ is 73.3.

Figure \ref{Fig4}(b) shows the normalized maximum growth thickness \textit{vs.}\
mask thickness at $t = T_3$ for contact angles $\theta > 90^\circ$.
The graphs for contact angles $\theta < 90^\circ$ are not shown since,
as Figure \ref{Fig2}(a) suggests, the maximum elevation of the surface
can occur at the contact point as well as at the ``bump" - thus sometimes making it 
difficult to distinguish between these two cases. The normalized maximum growth 
rate decreases with the mask thickness because the ``bump" smooths out
as the crystal grows vertically.  
The value of $T_3$ depends on 
the mask thickness; for the case of 
Figure \ref{Fig2}(b), 
$T_3 = 9.5$.

The numerical experiments described in this section and in Section \ref{Sec5}
were run on a single 300 MHz R12000 CPU on an 
SGI Origin 2000 Workstation. The computational
times depend strongly on $h_m, L$ and $\ell$ but they never exceeded 1 hour.

\Section{Numerical results for mask layer of zero thickness}

\label{Sec5}

In this section, we set $\bar h_m$ to zero in the boundary conditions
\rf{2.31}, \rf{2.34} and again solve the resulting initial/boundary
value problem for the evolution equations \rf{2.27} numerically.

\begin{figure}[h]
\centering
\psfig{figure=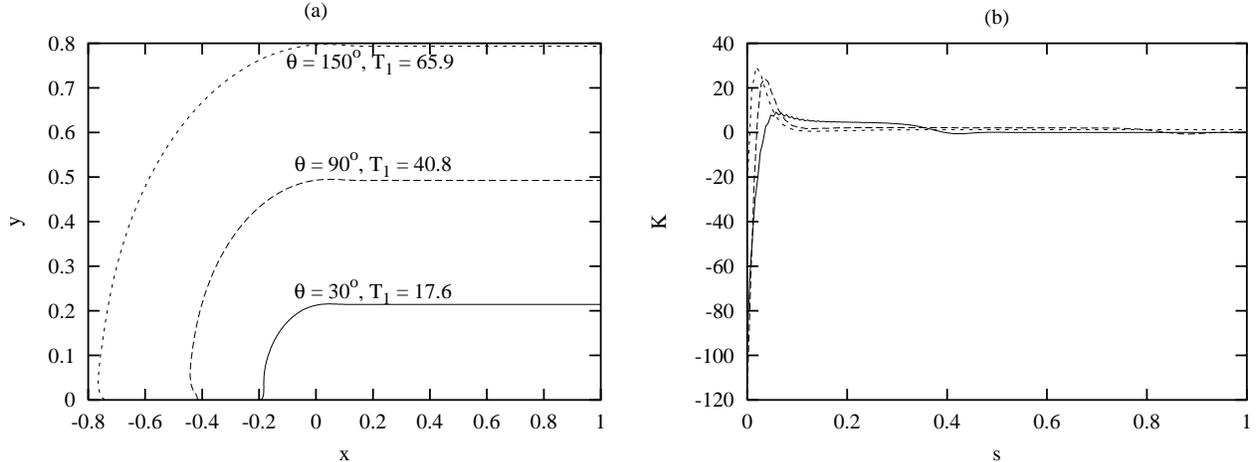,height=2.4in,width=6.5in,angle=270}
\caption{
(a): Graphs of the crystal surface at times the ``bump" is considered to
have disappeared, for different contact angles.
$L=\ell=5\ \mu \mbox{m},\ h_m = 0$.
(b): Curvatures of the curves in (a) as functions of the arc length.}
\label{Fig5}
\end{figure}

Figure \ref{Fig5}(a) shows the example of the crystal surface at time $T_1$
for different contact angles. The corresponding curvatures as functions
of the arc length are shown
in Figure \ref{Fig5}(b). The spikes on the graphs of $K(s)$ correspond to the
regions in the vicinity of the contact point on the graphs of the surface
where curves are bent to satisfy the equilibrium contact angle boundary
condition. The inflection points located at $x \approx 0.08,\ 0.11,\ 0.12$
(immediately after the smoothed ``bump") on the graphs of the surface
for $\theta=30^\circ,\ 90^\circ,\ 150^\circ$ respectively correspond
to $s \approx 0.39,\ 0.84,\ 1.33$ (the last point not shown) on the graphs
of $K(s)$.

\begin{figure}[h]
\centering
\psfig{figure=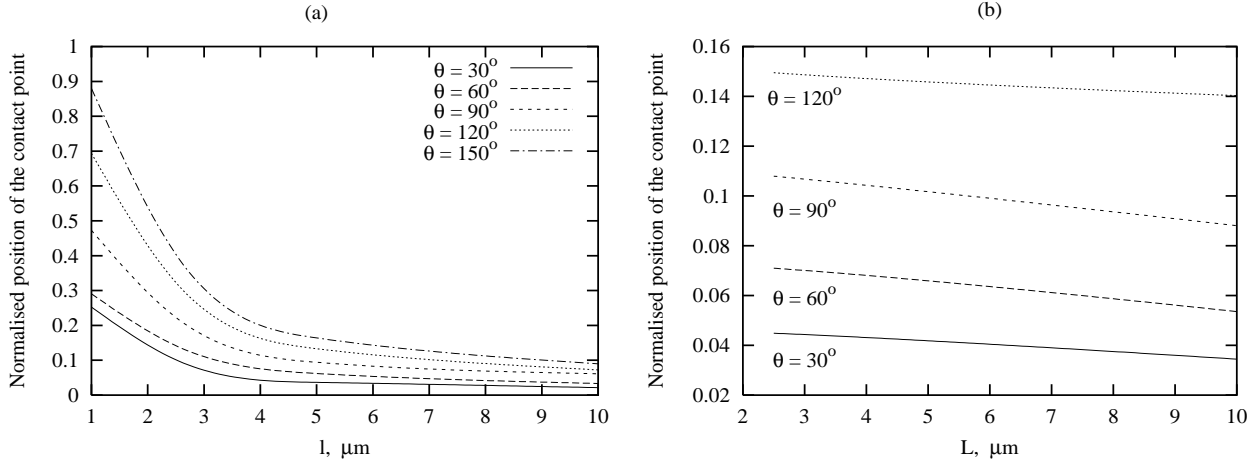,height=2.4in,width=6.5in,angle=270}
\caption{
(a): Normalized position at $t=T_1$
of the contact point \textit{vs.}\ mask region width, for different contact angles.
$L=5\ \mu \mbox{m},\ h_m = 0$.
(b): Normalized position (at $t=T_1$)
of the contact point \textit{vs.}\ substrate region width, for different contact angles.
$\ell=5\ \mu \mbox{m},\ h_m = 0$. For the representative values of $T_1$ refer to
Figure \ref{Fig5}(a).}
\label{Fig6}
\end{figure}

\begin{figure}[h]
\centering
\psfig{figure=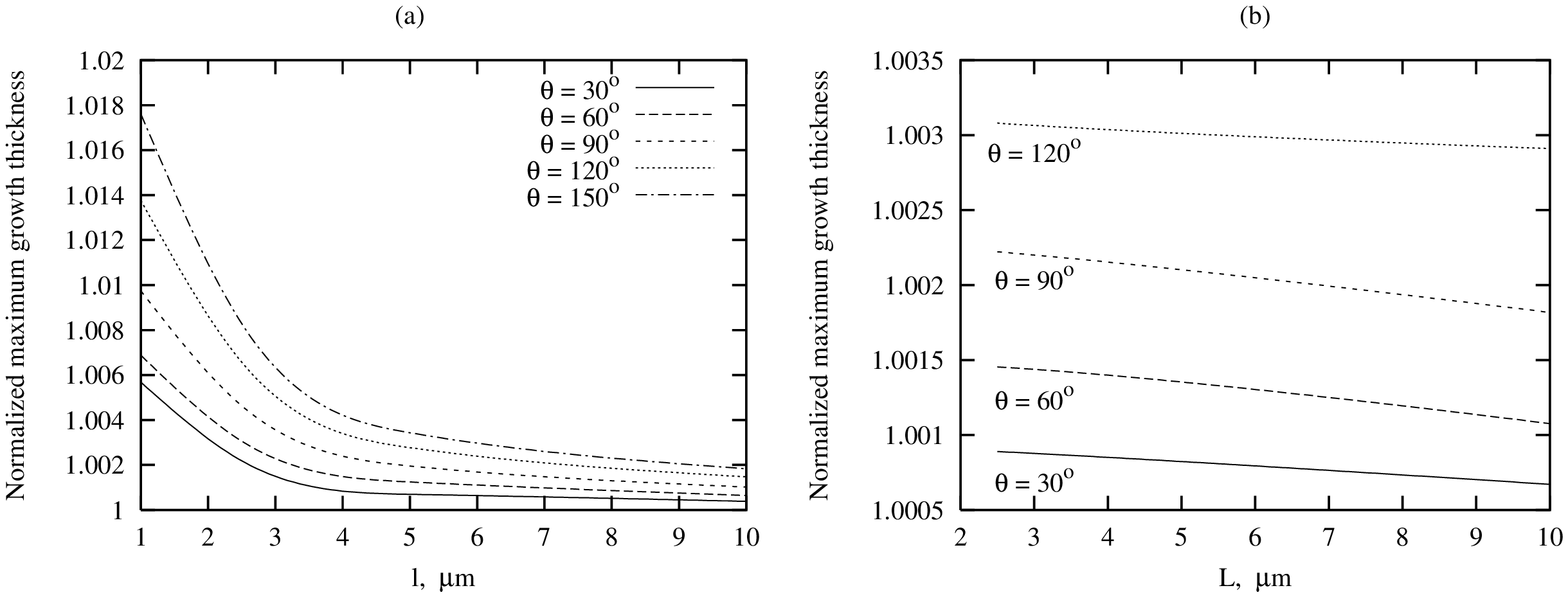,height=2.4in,width=6.5in,angle=270}
\caption{
(a): Normalized maximum growth
thickness (at $t=T_2$) \textit{vs.}\ mask region width, for different contact angles.
$L=5\ \mu \mbox{m},\ h_m = 0$.
(b): Normalized maximum growth
thickness (at $t=T_2$) \textit{vs.}\ substrate region width, for different contact angles.
$\ell=5\ \mu \mbox{m},\ h_m = 0$.}
\label{Fig7}
\end{figure}

In Figures \ref{Fig6}, \ref{Fig7} we study the dependence of the 
ELOG characteristics on widths of the mask and the substrate regions, $\ell$ and $L$.

Figure \ref{Fig6} shows the dependence of the normalized position
of the contact point on $\ell$ and $L$. As in the case of the mask layer
of nonvanishing thickness, the overgrowth onto the mask is larger for larger
contact angles. For a given contact angle, the overgrowth decreases with increasing
$\ell$ and $L$, with the dependence of the overgrowth in Figure \ref{Fig6}(b)
on $L$ being approximately linear.
Note that the variation in $\ell$ has stronger impact
on the overgrowth than the variation in $L$. Figure \ref{Fig6}(a) suggests
that wider mask regions will result in the saturation of the overgrowth.
This conclusion is consistent with the model. Indeed, the surface diffusion 
flux from the mask 
onto the growing crystal is the driving force for ELOG, with another component
of the driving force being the constant flux of atoms from the vapor on the
crystal surface. The flux on the mask is given by \rf{2.26} and it saturates as 
$d=\ell/L$ increases (the close-to-linear dependence of the
ELOG on $L$ can't be easily explained since all nondimensional parameters
of the problem are functions of $L$, ref. Table \ref{Table3}).

Figure \ref{Fig7} shows the similar dependence of the normalized maximum growth
thickness on $\ell$ and $L$ and the same explanation applies to the curves in Figure
\ref{Fig7}(a). It is interesting that the characteristic time 
$T_2$ does not depend on the
contact angle. That is, having $L$ fixed, the time needed for ELOG
over the distance $\ell$ is independent on the ratio of the crystal 
to the mask surface energies (for the case of 
Figure \ref{Fig5}(a), 
$T_2 = 87$).

\begin{figure}[h]
\centering
\psfig{figure=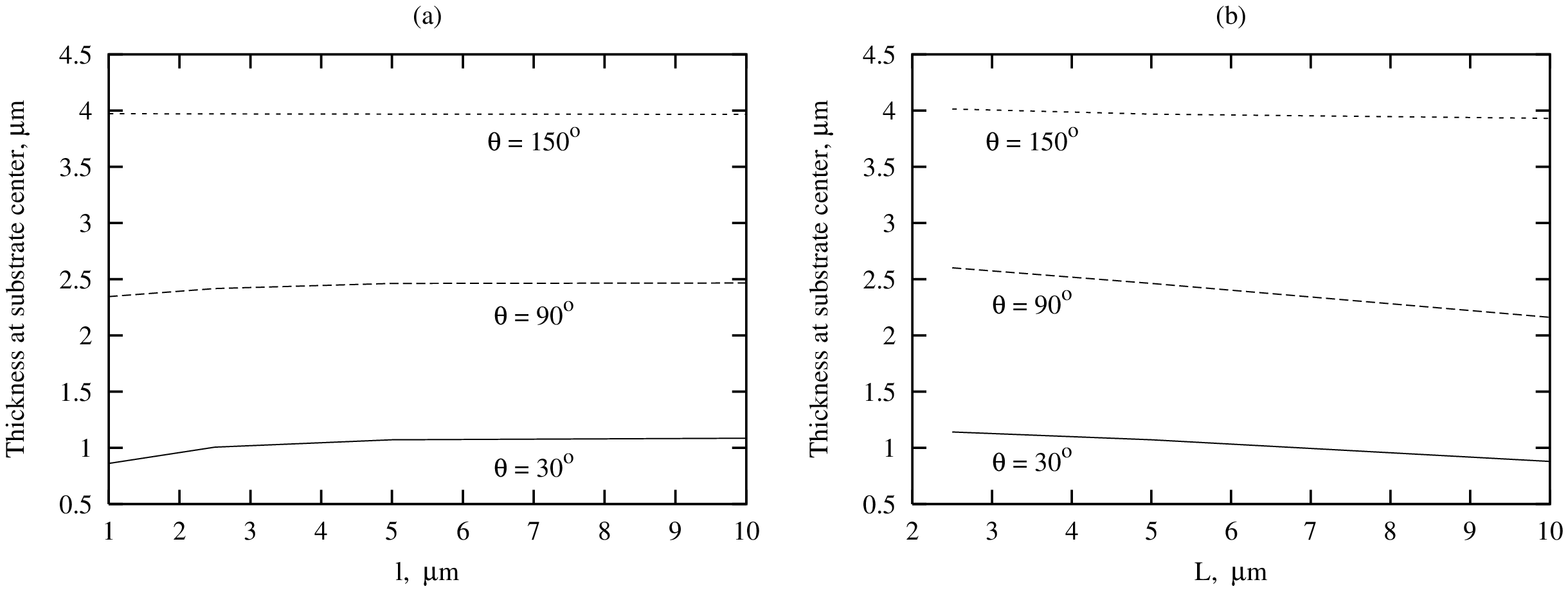,height=2.4in,width=6.5in,angle=270}
\caption{
(a): Thickness at substrate center
(at $t=T_1$, ref. Figure \ref{Fig5}(a)) \textit{vs.}\ mask region width, for different contact angles.
$L=5\ \mu \mbox{m},\ h_m = 0$.
(b): Thickness at substrate center
at $t=T_1$ \textit{vs.}\ substrate region width, for different contact angles.
$\ell=5\ \mu \mbox{m},\ h_m = 0$.}
\label{Fig8}
\end{figure}

Film thickness at the center of the substrate region ($x=L$) as a function of
$\ell$ and $L$ is shown in Figure \ref{Fig8}.  Film thickness increases with 
$\ell$, and the rate of the increase decreases as the contact angle increases
(Figure \ref{Fig8}(a)). Film thickness linearly decreases with $L$ 
(also with decreasing rate with respect to the increase of the contact
angle, Figure \ref{Fig8}(b)).

%

Film thickness at the center of the substrate region 
increases essentially linear in $t'$, with rate $\approx 1\ \mu \mbox{m}/\mbox{min}$.
There is no dependence of the growth at substrate center on the contact angle.

\section{Discussion and conclusions} 

\label{Sec6}

We developed a model for the epitaxial semiconductor crystal 
growth on a masked substrate in a surface diffusion-limited growth regime. 

In this model, the overgrowth of the crystal on the mask (as well as
growth in vertical direction) is a dynamic process which depends on many
factors, such as the height of the mask above the substrate,
the widths of masked and open regions, the ratio of mask to substrate
surface tensions, etc.  Unlike Type 1 and Type 2 models cited in the 
Introduction, our
model does not assume the overgrowth or lack of the overgrowth {\em a priori},
while in these models the assumption is made based on results of specific 
experiment which the model aims to describe.  Despite the fact that numerical 
results presented in Sections \ref{Sec4} and \ref{Sec5} always exhibit the overgrowth,
the trial runs with anisotropy included in the model indicated
that the overgrowth could be suppressed by selecting the degree of anisotropy 
and the fast growth directions \cite{FUTURE}. 

The predictions of the modeling compare favorably to the results obtained
by other authors. The region of enhanced growth near the contact point 
(``bump")
shown in Figure \ref{Fig2} supports the experimental and theoretical 
evidence for the surface diffusion on the mask to contribute strongly to such
enhanced growth \cite{GIBBON,SASAKI,SAKATA}, \cite{GREENSPAN}-\cite{ZYBURA}.
Eventually, the ``bump" is smoothed out by the surface diffusion on
the crystal surface.

The profiles of the crystal surface
we obtained  
show close resemblance to experimental profile in
Figure 5(b) of
\cite{THRUSH} and to experimental profiles in Figures 4 and 5 of \cite{SASAKI}. 
To make such comparison, our computed profiles should
be truncated, at some given time, with imaginary vertical
line passing to the left of the ``bump" maxima. The right part of the profile 
is then shifted to 
the right so that the left endpoint stays on the vertical 
line at $x=0$ (to account for the lack of the overgrowth in the works cited).
The detailed comparison with these and other experimental works is,
however, impossible since the natural anisotropy of the crystal growth
is not accounted for in this study.

The linear dependence of the normalized maximum growth thickness and of
the film thickness at the center of the substrate region  on the 
width of the substrate region $L$
(ref. Figures \ref{Fig7}(b), \ref{Fig8}(b)) was found experimentally 
in \cite{SASAKI} for the surface diffusion dominated growth.
Also, Figure \ref{Fig8}(a) shows that the film thickness at the center of the 
substrate region monotonically increases with the mask width $\ell$ due to
the increase of the lateral source supply . This result is well
known \cite{THRUSH,SAKATA,GREENSPAN}. Again, similar dependence
was found experimentally in \cite{SAKATA} for the surface diffusion dominated 
growth. Note that the normalized maximum growth thickness in Figure 
\ref{Fig7}(a) decreases with $\ell$ despite that the film thickness at $x=L$
increases with $\ell$. The reason is that surface diffusion
along the crystal surface has a strong smoothing effect in the regions
of the surface where the curvature is large.

Future directions for the research include considering anisotropic growth,
the diffusion in the vapor and the application of external fields.
The first of these results is already in hand, and will be reported
elsewhere \cite{FUTURE}.

\bigskip
\bigskip
\noindent
{\bf Acknowledgements}
\vspace{0.5cm}\\
\noindent M. Khenner and R.J. Braun were supported by NSF Grants DMS-9631287
and DMS-9722854. 
This problem
is an outgrowth of one studied in the 15th Annual Mathematical Problems in
Industry Workshop held in 1999.
The authors thank G. Berensel, 
K. Brattkus, L.P. Cook, A.D. Fitt, D.A. French, 
S. Kim, J.R. King, A.A. Lacey, J. Pelesko, D. Petrasek, C. Please
and A. Roosen for helpful conversations.
M.G. Mauk wishes to acknowledge support of the 
Ballistic Missile Defense Organization under 
Small Business Innovation Research contract 
DASG60-92-C-004.

\end{document}